\documentclass[sigconf,screen]{acmart}

\AtBeginDocument{%
  \providecommand\BibTeX{{%
    Bib\TeX}}}

\usepackage{hyperref}
\usepackage{comment}

\usepackage{amssymb}
\usepackage{bbding}
\usepackage[linesnumbered,ruled,vlined]{algorithm2e}
\SetAlgoNlRelativeSize{0}
\SetNlSty{textbf}{}{}
\usepackage{textcomp}
\usepackage{xcolor}
\usepackage{enumitem}
\usepackage{balance}
\usepackage{subfigure}
\usepackage{multirow}
\usepackage{mdframed}
\usepackage{tcolorbox}
\usepackage[]{collab}
\usepackage{subcaption}
\usepackage{makecell} 
\usepackage{graphicx}
\usepackage{ulem}
\usepackage{algpseudocode}
\usepackage{amsmath}
\usepackage{tabularx}    
\usepackage{multirow}    
\usepackage{booktabs}    
\usepackage{array}  

\usepackage[font=normalsize,labelfont=bf,format=plain]{caption}

\usepackage{tikz}
\usepackage{collab}
\usepackage{setspace} 
 
\acmConference[ICSE '26]{}{Apr 14 -- 21,
  2026}{Rio De Janeiro, Brazil}
  
\renewcommand\footnotetextcopyrightpermission[1]{} 
\collabAuthor{YL}{orange}{Alan}

\def\BibTeX{{\rm B\kern-.05em{\sc i\kern-.025em b}\kern-.08em
    T\kern-.1667em\lower.7ex\hbox{E}\kern-.125emX}}

\newcommand{\ie}{{\textit{i.e.}},\xspace}
\newcommand{\eg}{{\textit{e.g.}},\xspace}

\definecolor{ballblue}{rgb}{0.13, 0.67, 0.8}
\definecolor{OLGreen}{rgb}{0.239, 0.522, 0.290}
\collabAuthor{yc}{blue}{Yichen Li}
\collabAuthor{yl}{OLGreen}{Yulun Wu}
\collabAuthor{gb}{orange}{Guangba Yu}

\definecolor{mygray}{RGB}{211,211,211} 
\definecolor{morandi}{RGB}{210,225,225} 

\definecolor{myyellow}{HTML}{FFF2CC}
\newcounter{insight}
\newcommand{\insight}[1]{\refstepcounter{insight}
	\begin{mdframed}[linecolor=gray!25,roundcorner=12pt,backgroundcolor=myyellow!20,linewidth=3pt,innerleftmargin=2pt, leftmargin=0cm,rightmargin=0cm,topline=false,bottomline=false,rightline=false,leftline=false]
		\textbf{Insight \arabic{insight}:} #1
	\end{mdframed}
}

\newcommand{\method}{Autoscope}

\newcommand{\find}[1]{
    \begin{tcolorbox}[colback=black!5!white,boxrule=0pt,top=0pt,bottom=0pt,left=0pt,right=0pt,boxsep=1.1mm,nobeforeafter]
        \emph{#1}
    \end{tcolorbox}
}

\begin{document}

\title{Trace Sampling 2.0: Code Knowledge Enhanced Span-level Sampling for Distributed Tracing}

\author{Yulun Wu}
\affiliation{%
  \institution{The Chinese University of Hong Kong}
  \city{Hong Kong SAR}
  \country{China}
}

\author{Guangba Yu}
\authornote{Corresponding author.}
\affiliation{%
  \institution{The Chinese University of Hong Kong}
  \city{Hong Kong SAR}
  \country{China}
}

\author{Zhihan Jiang}
\affiliation{%
  \institution{The Chinese University of Hong Kong}
  \city{Hong Kong SAR}
  \country{China}
}

\author{Yichen Li}
\affiliation{%
  \institution{The Chinese University of Hong Kong}
  \city{Hong Kong SAR}
  \country{China}
}

\author{Michael R. Lyu}
\affiliation{%
  \institution{The Chinese University of Hong Kong}
  \city{Hong Kong SAR}
  \country{China}
}


\renewcommand{\shortauthors}{Trovato et al.}




\begin{abstract}
Distributed tracing is an essential diagnostic tool in microservice systems, but the sheer volume of traces places a significant burden on backend storage. A common approach to mitigating this issue is trace sampling, which selectively retains traces based on specific criteria, often preserving only anomalous ones. However, this method frequently discards valuable information, including normal traces that are essential for comparative analysis.  To address this limitation, we introduce Trace Sampling 2.0, which operates at the span level while maintaining trace structure consistency. This approach allows for the retention of all traces while significantly reducing storage overhead. Based on this concept, we design and implement Autoscope, a span-level sampling method that leverages static analysis to extract execution logic, ensuring that critical spans are preserved without compromising structural integrity.  

We evaluated Autoscope on two open-source microservices. Our results show that it reduces trace size by 81.2\% while maintaining 98.1\% faulty span coverage—outperforming existing trace-level sampling methods. Furthermore, we demonstrate its effectiveness in root cause analysis, achieving an average improvement of 8.3\%. These findings indicate that Autoscope can significantly enhance observability and storage efficiency in microservices, offering a robust solution for performance monitoring.
\end{abstract}


\maketitle

\section{Introduction}
\begin{figure*}[t]
\centering
\includegraphics[width=\textwidth]{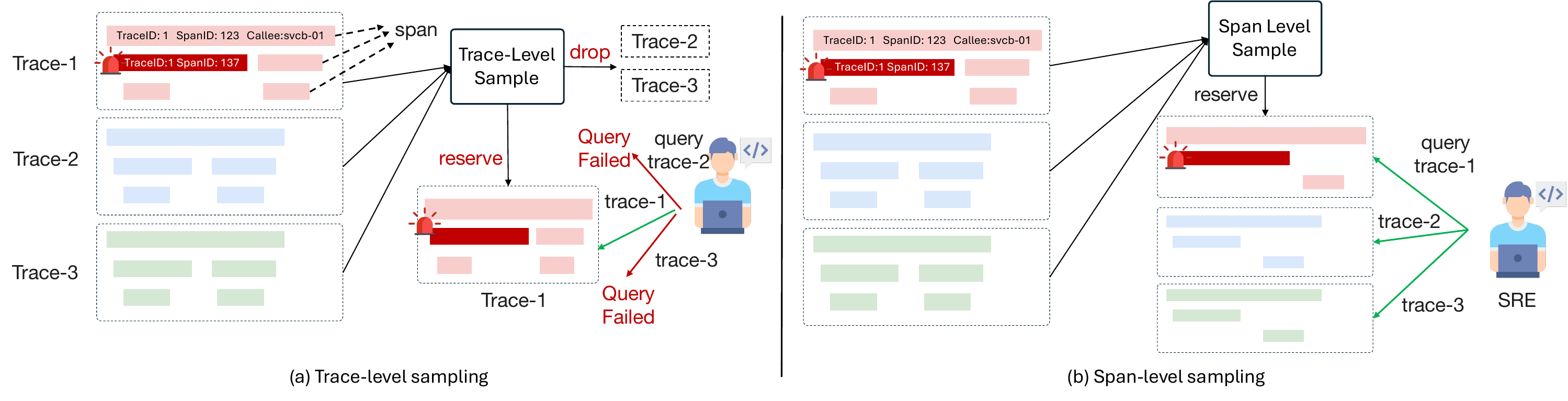} 
\vspace{-0.30in}
\caption{Comparisons between trace- and span-level sampling.}
\vspace{-0.1in}
\label{fig:span-sample}
\end{figure*}

Modern software architectures have evolved into distributed microservice systems~\cite{newman2021building,tse2019zhou}, making distributed tracing an essential observability tool for understanding application behavior and performance at scale~\cite{sigelman2010dapper,Mint2025Asplos,Fonseca2007Xtrace,liu2023prism}. 
As shown the \emph{Trace-1} in Fig.~\ref{fig:span-sample}, by capturing fine-grained spans along request paths across services, traces create a comprehensive execution path that includes service transitions, time distributions, and context propagation. These tracing systems enable Site Reliability Engineers (SREs) to effectively profile system performance~\cite{Zhang2022CRISP,Huang2021tprof,Zhou2019fse}, identify performance anomalies~\cite{Doray2017TraceCompare,Xu2021TraceLingo,li2020traceanomaly}, and diagnose root causes~\cite{Murali2021Minesweeper,li2021traceRCA,yu2021microrank}. Therefore, many industry leaders have embraced various tracing solutions, including OpenTelemetry~\cite{Opentelemetry}, Skywalking~\cite{Skywalking}, and Jaeger~\cite{Jaeger}, demonstrating the technology's widespread adoption~\cite{guo2020GMTA,cai19traceali, Zhang2022CRISP}.

Although distributed traces offer valuable information for system analysis, their extensive volume and the resulting storage requirements create significant obstacles. Alibaba’s e-commerce platform generates between 18.6 and 20.5 pebibytes (PB) of trace data each day~\cite{Mint2025Asplos}. This immense amount arises because developers seek to capture the full spectrum of application behaviors to facilitate the diagnosis of emerging issues. However, persistently storing such a large amount of trace data incurs substantial operational overhead. To address these challenges, trace-level sampling techniques~\cite{Gias2023SampleHST, zhang2023benefit, Pedro2019sifter, huang2021sieve, TraStrainer2024Huang} have been devised to selectively retain traces relevant to anomalous behavior~\cite{yu2021microrank,Doray2017TraceCompare,li2020traceanomaly}.

In this paper, we designate the trace-level sampling as ``trace sampling 1.0''. As shown in Fig.~\ref{fig:span-sample}-(a), the primary mechanism of trace-level sampling involves initially identifying the traces to be sampled and subsequently retaining only those selected traces while completely eliminating the remainder, a process we call the ``1 or 0'' strategy. These techniques are generally classified into head sampling~\cite{Kaldor2017Canopy,sigelman2010dapper} or tail sampling~\cite{Pedro2019sifter, huang2021sieve, TraStrainer2024Huang}, depending on when the sampling decision occurs and the criteria applied. However, this approach of wholly discarding unsampled traces reveals significant shortcomings in practical scenarios. Existing research~\cite{Mint2025Asplos} indicates that SREs may still need to access these discarded traces, as the traits of traces that require analysis are often unpredictable, leading to a query miss rate of up to $27.17\%$. Moreover, such trace query failures can substantially hinder diagnostic approaches based on comparing normal and abnormal traces~\cite{Doray2017TraceCompare,yu2021microrank,li2020traceanomaly}.

To address the limitations of trace-level sampling and enhance sampling flexibility, we introduce the concept of span-level sampling, which we designate as ``trace sampling 2.0''. The fundamental observation behind span-level sampling is that most spans within a trace are irrelevant to explaining the performance variations under scrutiny~\cite{Kaldor2017Canopy, tse2019zhou, ding2015log2}. For example, a study of Alibaba's trace data indicates that 90\% of spans contribute little meaningful information, while only 10\% are critical for diagnosing performance issues~\cite{AliTrace2021Socc}. This suggests the potential to balance trace cost and utility at the span level by preserving spans that aid fault diagnosis while discarding those that do not.

However, achieving a balance between trace cost and utility at the span level is far from straightforward. Transitioning from the trace-level sampling approach, an intuitive strategy might involve analyzing variations in span duration and retaining spans that significantly deviate from typical latency patterns. Nevertheless, we observed that relying solely on duration data overlooks critical invocation details within the trace, such as the specific services or functions traversed. For example, as illustrated in Figure~\ref{fig:motivation}-(a) and (b), \emph{Trace-1} and \emph{Trace-2} represent different request types, but after sampling, 
their trace structure becomes identical, which makes their request types indistinguishable, thus affecting downstream diagnostic tasks (§~\ref{sec:Motivating}).

To overcome this limitation, we propose  \method, the first span-level trace sampling method designed to advance Trace Sampling 2.0. The core idea of this work lies in precisely extracting execution logic from the intricate source code of a distributed system to enhance the span-level sampling process. Specifically, Autoscope begins by constructing a Call Site Control Flow Graph (CSCFG) through static analysis. Since static analysis alone struggles to capture cross-service calls in microservices architectures, Autoscope integrates runtime data for optimization. During the sampling phase, Autoscope maps span to their corresponding code functions and partitions trace data based on the CSCFG. This enables the identification of Dominant Span Sets (DSS), which leverages inference relationships between spans. By recording just one span within a DSS, the entire set can be inferred, significantly improving sampling efficiency.  To ensure the representativeness of sampled spans, Autoscope employs a robust Z-score anomaly detection method to quantify span anomalies and select spans at the DSS level. Additionally, Autoscope adopts an incremental path-matching strategy to optimize execution path matching, further enhancing the completeness and accuracy of sampled trace data.

We evaluated the performance of Autoscope on two open-source microservice applications. The results show that Autoscope effectively reduces trace size by 81.2\% using span-level sampling while preserving all request traces record, significantly improving sampling efficiency. Despite this reduction, the sampled traces maintain high quality, with faulty span coverage reaching 98.1\%, outperforming all trace-level sampling methods.  To further assess trace quality, we conducted experiments on Root Cause Analysis (RCA). Autoscope’s sampled traces consistently outperformed other sampling strategies across four SOTA RCA methods, achieving an average improvement of 8.3\%. These findings confirm the effectiveness and high quality of Autoscope’s sampling approach.

In summary, we make the following contributions in this paper:
\begin{itemize}[leftmargin=*, topsep=0pt, parsep=0pt]
    \item  We propose the concept of Trace Sampling 2.0, introducing a span-level approach that precisely identifies critical spans within traces while drastically reducing storage costs.
    \item We design and implement Autoscope to achieve Trace Sampling 2.0, a novel sampling method that maps spans to their functions based on Call Site Control Flow Graphs (CSCFGs). By leveraging static analysis, Autoscope identify Dominant Span Sets, preserving essential spans for both the trace structure and anomalies.
    \item We extensively evaluate Autoscope on two microservice systems, Autoscope achieves an 81.2\% reduction in trace data while maintaining a 98.1\% coverage of faulty spans, and demonstrates an obvious improvement on downstream RCA tasks, indicating its superior advantage over traditional sampling approaches.
\end{itemize}
\vspace{-5pt}
\section{Background}\label{sec: background}

\subsection{Distributed Tracing}
 Distributed tracing~\cite{sigelman2010dapper} captures causality information within the distributed environment, allowing it to be transmitted across process boundaries. This mechanism facilitates the inference of system states across diverse services and functions throughout the lifecycle of a request, thereby aiding in the identification of code regions responsible for performance bottlenecks~\cite{Huang2021tprof,Zhang2022CRISP}.

\begin{figure}[t]
\includegraphics[width=0.5\textwidth]{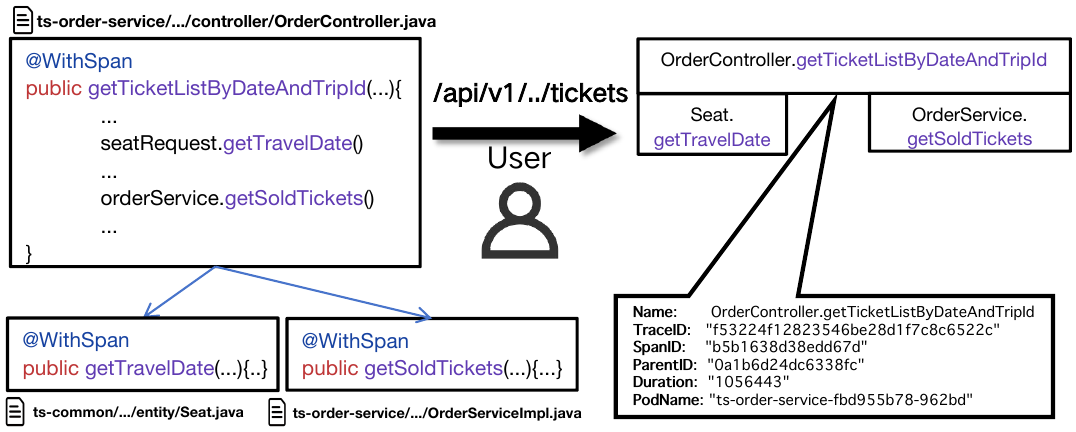} 
\vspace{-10pt}
\caption{Code and Distributed Tracing}
\vspace{-10pt}
\label{fig:BG_1}
\end{figure}

We use a real example from TrainTicket~\cite{TrainTicket}, a widely adopted open source microservices system, to illustrate the relationship between user code, span, and traces. In this example, we follow the manner of widely used trace framework OpenTelemetry~\cite{Opentelemetry}. As shown in Figure~\ref{fig:BG_1}, when a  user tries to trigger a request to a certain URL, it will then invoke 
\texttt{getTicketListByDateAndTripId()} along with its child functions, generating corresponding traces and spans. For clarity, some details are omitted in the diagram.

\textbf{Span.} A Span represents a request-response interaction, encapsulating API or function calls within a running service instance. As shown in the left of Fig.~\ref{fig:BG_1}, we annotate the \texttt{OrderService}'s function \texttt{getTicketListByDateAndTripId()}, along with its subfunctions \texttt{getTravelDate()} and \texttt{getSoldTickets()}, using \texttt{withSpan}\footnote{\url{https://opentelemetry.io/docs/zero-code/java/agent/annotations/}}. This annotation instructs OpenTelemetry to generate the corresponding spans. On the right side of the figure, these Spans are visualized as blocks, each containing metadata such as a unique span ID, start time, duration, and parent ID, which reflects the function call hierarchy. As the fundamental building blocks of distributed traces, spans represent discrete computational tasks in a distributed system.

\textbf{Trace.} A Trace consists of multiple spans, collectively representing the end-to-end execution of a request within a microservices system. In Fig.~\ref{fig:BG_1}, the trace comprises spans corresponding to three functions, mirroring the function call relationships. For instance, the Span \texttt{OrderController.getTicketListByDateAndTripId()} is the parent of \texttt{Seat.getTravelDate()}, meaning that the function \texttt{getTicketListByDateAndTripId()} invokes the \texttt{getTravelDate()}, forming a Caller-Callee relationship. Thus, the trace captures the execution flow of the static code, while the logical code structure of the application dictates the organization of the Trace and the execution order of Spans, the two are closely intertwined.

\vspace{-6pt}
\subsection{Call-Site Control Flow Graph}\label{sec:CSCFG}
\begin{figure}[t]
\includegraphics[width=0.5\textwidth]{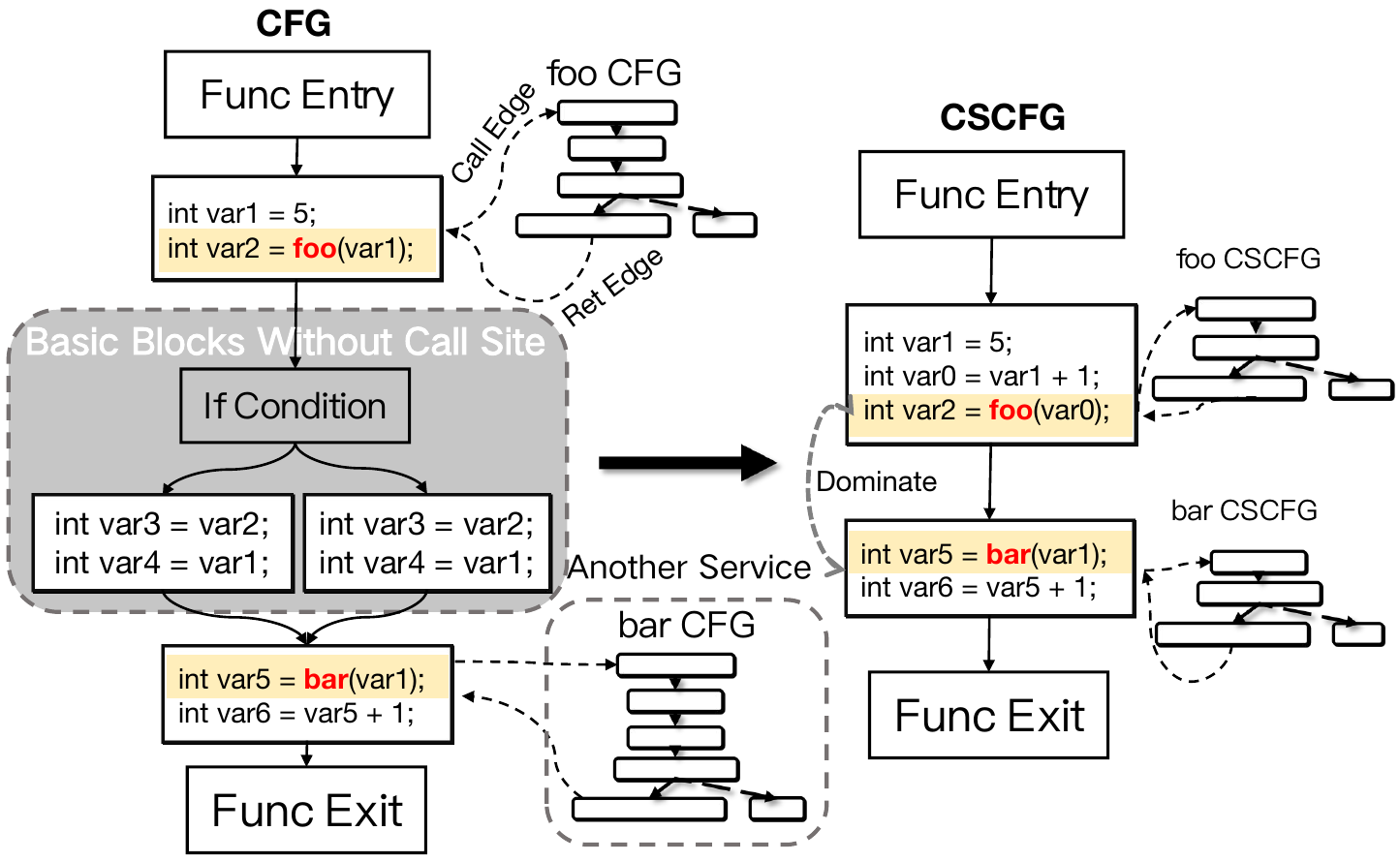} 
\caption{Examples of CFG and CSCFG}
\label{fig:CSCFG}
\end{figure}

Static code analysis examines source code or bytecode without executing the program, aiming to identify potential errors and enhance code comprehension~\cite{chess2004static}. The core of this approach lies in analyzing code structure, control flow, and other program properties using techniques such as pattern matching and abstract interpretation~\cite{TanL23ISSTA}. One of the most common representations in static analysis is the Control Flow Graph (CFG)~\cite{allen1970control}, which models the control flow within a function. As shown in Figure~\ref{fig:CSCFG}, a CFG consists of multiple Basic Blocks (BBs), each containing a sequence of instructions that execute in order. These BBs are connected by edges that represent the possible execution paths.

In a traditional CFG, function calls establish connections between different CFGs, forming an Inter-Procedural Control Flow Graph (ICFG)~\cite{nielson1999interprocedural}. For example, in Fig.~\ref{fig:CSCFG}, the CFGs of \texttt{foo()} and \texttt{bar()} are linked to the main function through \textit{Call Edges} and \textit{Return Edges}.
This kind of edge can also bridge the function across services under microservices environments, like the \texttt{bar()} function~\cite{COCAICSE25}.

However, in microservices, tracing works at the function level, which means traces primarily record function call relationships rather than the complete CFG structure. Therefore, this work focuses on a variant of CFG known as the Call-Site CFG (CSCFG)~\cite{wu2016casper}. Unlike traditional CFGs, CSCFG retains only BBs that contain function calls (as shown on the right side of Figure~\ref{fig:CSCFG}). This selective representation aligns more closely with the trace spans.

One key property of CSCFG is the \textit{Dominate Relation}~\cite{alfred2007compilers} between functions. In a CFG, if every path from function entry to exit must pass through basic block A before reaching B, then B is said to dominate A. Similarly, if A also dominates B in all backward paths from exit to entry, then A and B are mutually dominant. This relationship allows for function call inference when the execution flow is deterministic. For instance, in the example, since \texttt{foo()}  and \texttt{bar()}have a mutual dominance relationship, the presence of \texttt{foo()}  in the execution flow ensures the existence of \texttt{bar()}, and vice versa. The same principle applies to trace spans, where mutual dominance enables span call inference.

\vspace{-6pt}
\section{Trace Sampling 2.0: What, Why and How}\label{sec:Motivating}

\begin{figure*}[t]
\centering
\includegraphics[width=\textwidth]{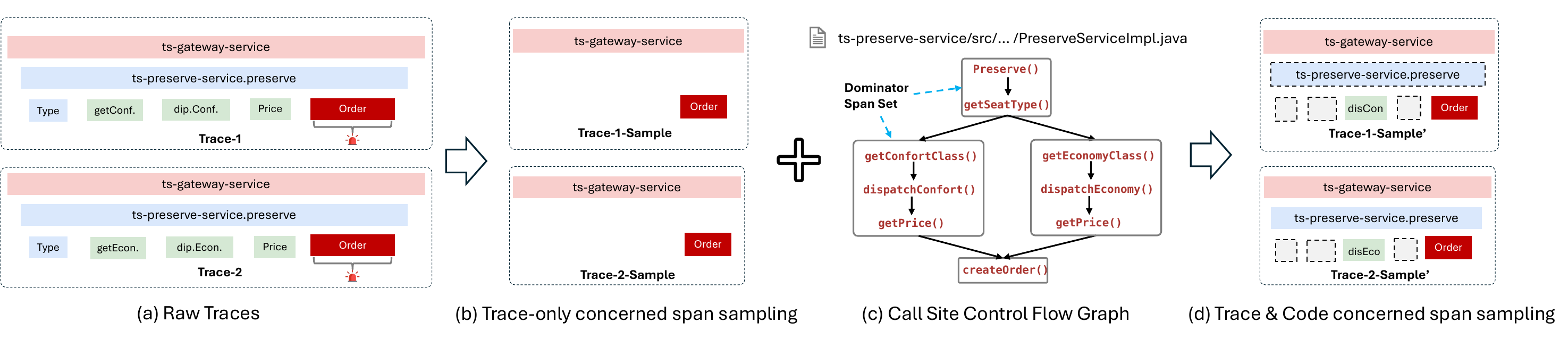} 
\vspace{-0.30in}
\caption{An example of the importance of source code in span-level sampling.}
\vspace{-0.1in}
\label{fig:motivation}
\end{figure*}

\textit{Trace Sampling} refers to the process of selecting a subset of traces from a massive number of traces according to a specified strategy, so that only some requests are recorded and preserved. This sampling mechanism is essential because capturing every request in production environments often leads to overwhelming and unsustainable storage and analysis costs. Today, Trace Sampling has been widely deployed in industry. A common sampling approach used by tracing frameworks such as Jaeger~\cite{Jaeger} and Opentelemetry~\cite{Opentelemetry} is uniform random sampling at a fixed rate (e.g., 5\%), whereby the system decides at the beginning of a request whether to record the trace. This approach, often referred to as \textit{head-based sampling}, does not consider the varying analytical value of individual traces~\cite{sigelman2010dapper,Kaldor2017Canopy}.
To address this limitation, prior studies have proposed and adopted \textit{tail-based sampling}~\cite{Pedro2019sifter, huang2021sieve, TraStrainer2024Huang}, which makes the sampling decision at the termination of a request. By leveraging the complete trace information, tail-based sampling can better capture traces with higher diagnostic value. 

In this paper,  both head-based and tail-based approaches are classified as \textit{trace-level sampling} strategies, employing a ‘1 or 0’ strategy: any trace flagged for sampling is fully retained, while all other traces are discarded. We refer to these \textit{trace-level sampling} techniques collectively as \textit{Trace Sampling 1.0}. To offer more flexible sampling strategies, we introduce the concept of \textit{Trace Sampling 2.0}, which evolves from the original \textbf{trace-level sampling} (i.e., the ‘1 or 0’ strategy) to a more granular \textbf{span-level sampling} (i.e., Trace Sampling 2.0). Figure~\ref{fig:span-sample} shows a comparison between trace- and span-level sampling.  The remainder of this section presents the concept, motivation, and implementation of Trace Sampling 2.0.

\subsection{What is Trace Sampling 2.0?}

Trace Sampling 2.0 is a flexible sampling strategy that operates within a single distributed trace, selecting and retaining specific spans based on their significance. Compared to Trace Sampling 1.0, which determines whether to keep or discard an entire trace in its entirety, Trace Sampling 2.0 focuses on preserving critical spans (e.g., those with certain error codes, high response times, or on crucial business paths). With this approach, each trace remains present in the system, but only the most critical and relevant segments are stored. For SREs in need of diagnostic information, Trace Sampling 2.0 can substantially reduce the volume of data while still enabling rapid isolation of essential span segments. Consequently, this method maximizes visibility into key operations and supports more focused debugging and maintenance.

\begin{table}[t]
\caption{Comparison of SOTA trace sampling approaches.}
\vspace{-0.1in}
\resizebox{\columnwidth}{!}{
\begin{tabular}{@{}clccccc@{}}
\toprule
\multicolumn{2}{c}{\multirow{2}{*}{Method}}                                                        & \multicolumn{3}{c}{Input} & \multicolumn{2}{c}{Sampling Result} \\ \cmidrule(l){3-7} 
\multicolumn{2}{l}{}                                                                         & Trace   & Metric  & Code  & Normal Trace     & Abnormal Trace   \\ \midrule
\multirow{5}{*}{\begin{tabular}[c]{@{}l@{}}Trace-Level \\ Sampling\end{tabular}} & Random(5\%)~\cite{Jaeger}    & \textbf{\CheckmarkBold}        &         &       & Partial          & Negligible          \\
 & Perch~\cite{Pedro2018PERCH}       & \textbf{\CheckmarkBold} &  &  & Negligible & Comprehensive \\
 & Sifter~\cite{Pedro2019sifter}      & \textbf{\CheckmarkBold}  &  &  & Negligible & Comprehensive \\
 & TraceMesh~\cite{Chen2024Tracemesh}   & \textbf{\CheckmarkBold} &  &  & Negligible & Comprehensive \\
 & TraStrainer~\cite{TraStrainer2024Huang} & \textbf{\CheckmarkBold} & \textbf{\CheckmarkBold} &  & Partial    & Comprehensive \\ \midrule
\begin{tabular}[c]{@{}l@{}}Span-Level \\ Sampling\end{tabular}                   & Autoscope(Ours) & \textbf{\CheckmarkBold}          &        &  \textbf{\CheckmarkBold}     & Comprehensive    & Comprehensive    \\ \bottomrule
\end{tabular}%
}
\raggedright
\footnotesize{Note: Negligible (<1\% of normal $/$ anomalous traces), Partial (1\%-20\%), Substantial (20\%-80\%), Comprehensive (>80\%).}
\label{tab:sampling_comparison}
\end{table}

\subsection{Why do We Need Trace Sampling 2.0?}
\label{sec:why_we_need}
Figure~\ref{fig:span-sample}-(a) provides an example of trace-level sampling. The figure includes three traces, where \textit{Trace-1} contains \textit{Span-137} showing a performance issue, while \textit{Trace-2} and \textit{Trace-3} exhibit normal latency. Under conventional tail-based sampling~\cite{Pedro2019sifter, huang2021sieve, TraStrainer2024Huang}, only \textit{Trace-1} would be retained, and \textit{Trace-2} and \textit{Trace-3} would be discarded. However, in certain fault diagnosis scenarios(\eg off-the-path problem\cite{wu2019zeno}), SREs may wish to compare \textit{Trace-2} with \textit{Trace-1} to identify the potential root cause of the observed performance anomaly. Since \textit{Trace-2} has already been discarded, the diagnosis process is hampered, leading to inefficient troubleshooting.

Table~\ref{tab:sampling_comparison} quantifies the retention capabilities of various state-of-the-art trace sampling approaches, highlighting their effectiveness in preserving normal and abnormal traces. Random sampling, due to its inherent randomness, retains only a partial fraction (1\%--20\%) of normal and abnormal traces, limiting its utility for comprehensive analysis. Tail-based methods, including Perch~\cite{Pedro2018PERCH} , Sifter~\cite{Pedro2019sifter}, TraceMesh~\cite{Chen2024Tracemesh}, excel at capturing most abnormal traces (coverage $>80\%$), yet they retain negligible amounts ($<1\%$) of normal traces. This skewed retention leads to frequent query misses when SREs attempt to access normal traces for diagnostic purposes. Existing research underscores the importance of these normal traces, noting that in industrial systems, query miss rates can reach as high as 27.17\%~\cite{Mint2025Asplos}. Such a significant miss rate is non-trivial, as it directly undermines the ability to perform thorough root-cause analysis, motivating the need for sampling strategies that balance the retention of both trace types.

In contrast, Figure~\ref{fig:span-sample}-(b) illustrates the outcome of applying span-level sampling to the same traces. By storing approximately the same number of spans yet retaining a greater number of traces, this approach ensures that users can still retrieve the main structure of each trace along with its critical spans. Therefore, span-level sampling can increase the likelihood of successful queries, which is an essential aspect of operational troubleshooting. In general, span-level sampling introduces a novel perspective for balancing trace cost and utility. Its key advantages include:
\begin{itemize}[leftmargin=*, topsep=0pt]
\item \textbf{Enhanced flexibility:} By filtering at the span level, the system can selectively retain only the critical paths or the relevant spans of anomalies.
\item \textbf{Improved queryability:} Even with high sampling rates, each trace still retains essential information, reducing cases where entire traces are discarded and consequently become unavailable for querying.
\item \textbf{Controlled storage overhead:} Through informed decisions on which spans to preserve, this strategy helps to maintain a relatively comprehensive view of the system while remaining within acceptable storage limits.
\end{itemize}

\insight{Traditional trace-level sampling discards normal traces while retaining only anomalous ones, limiting trace queryability and fault diagnosis. In contrast, span-level sampling retains critical spans across more traces, enhancing cost-utility balance, thus motivating a shift to finer-grained sampling strategies.}

\subsection{How Can We Implement Trace Sampling 2.0?}
Implementing Trace Sampling 2.0 entails transitioning from trace-wide decisions to span-specific strategies. An intuitive approach, once moving away from conventional trace-level sampling, is to analyze spans for significant deviations in latency and retain only those exhibiting abnormal performance. For example, Figure~\ref{fig:motivation}-(a) shows two requests in the \textit{Trainticket} benchmark. The ``preserve'' request is triggered when the user interacts with the front-end UI, which then invokes the corresponding URL endpoint. Although both requests call the \texttt{ts-preserve-service}, one user is purchasing a comfortable class seat, whereas another is buying an economic class seat, leading to different underlying processing logic. In this scenario, \texttt{ts-preserve-service} encounters an issue during the execution of \texttt{createOrder()}, resulting in excessive latency.

If span-level sampling is based solely on trace latency, the outcome would resemble the sampling result shown in Figure~\ref{fig:motivation}-(b). Although both requests preserve the \texttt{createOrder()} span, an SRE still faces ambiguity. Specifically, the SRE needs to distinguish whether the failure occurred during comfortable or economic train seat purchase, but this method fails to capture such contextual differences between the two traces. Consequently, effective fault analysis may be hindered by the ambiguity introduced when sampling decisions consider only latency deviations.

\insight{
Span-level sampling based solely on latency deviations leads to ambiguity for SREs in pinpointing failure causes, motivating the need for Trace Sampling 2.0 to incorporate span-specific strategies that capture both performance anomalies and contextual nuances for effective fault analysis.
}

However, relying solely on the limited trace information fails to capture the granular and domain-specific contextual nuances that can drastically affect how a system behaves under various conditions.  In the ``comfortable class seat vs. economic class seat'' example, distinguishing the root cause of latency or failure requires more than simply knowing that a particular service method (\texttt{createOrder()}) is slow. SREs must understand which portion of the code logic is exercised, whether it is for a first-class seat purchase path or for a second-class seat purchase path, and how these two paths differ in their internal computations, such as additional validation steps for premium bookings or simpler queries for standard ones.

This ambiguity arises primarily from the inability of trace data to fully represent a program’s critical paths. Traces are inherently tied to user request types, and given the variability in user behavior, they often lack coverage of all possible request scenarios. For instance, if users predominantly request second-class seats, the trace may omit the execution path for first-class purchases, leaving SREs blind to potential bottlenecks in that logic. Consequently, achieving a comprehensive view of a program’s critical paths solely through traces remains elusive.

The evolution of code analysis tools~\cite{TanL23ISSTA, shi2018pinpoint} and the rise of open-source software present a compelling opportunity to address this gap. With application source code increasingly accessible to SRE, either from internal repositories or public platforms, static analysis of the codebase emerges as a viable means to extract critical paths. As shown in Fig.~\ref{fig:motivation}-(c), by mapping code's control flow logic to potential traces, such as distinguishing the conditional branches for comfortable versus economic seat processing in \texttt{createOrder()}, span-level sampling can be augmented with a priori knowledge of all possible paths. This enriched context enables span-level sampling to preserve critical path to avoid ambiguity.

If the code control flow logic of ``preserve'' is available(\ie like Fig.~\ref{fig:motivation}-(c) ), we can discern that ``comfort'' and ``economy'' classes occupy two distinct execution paths. Hence, informative spans such as \texttt{getComfortClass()}, \texttt{dispatchComfort()}, and \texttt{getPrice()} must be preserved to mark a request as targeting the comfort class. As illustrated in  Fig.~\ref{fig:motivation}-(d), retaining this critical branch yields a refined trace sampling result that resolves earlier ambiguities. Furthermore, it can further reconstruct the complete trace structure based on this code logic, thereby mitigating the impact of trace sampling on subsequent queries and diagnostic tasks. This observation underscores the potential of leveraging code insights to enhance span-level sampling strategies.

\insight{Integrating static code analysis enables span-level sampling to preserve critical branches, resolving ambiguity and enhancing fault diagnosis. This inspires a code-enhanced approach to improve span-level sampling effectiveness.}

\section{\method}\label{sec: method}

\begin{figure}[t]
\includegraphics[width=0.45\textwidth]{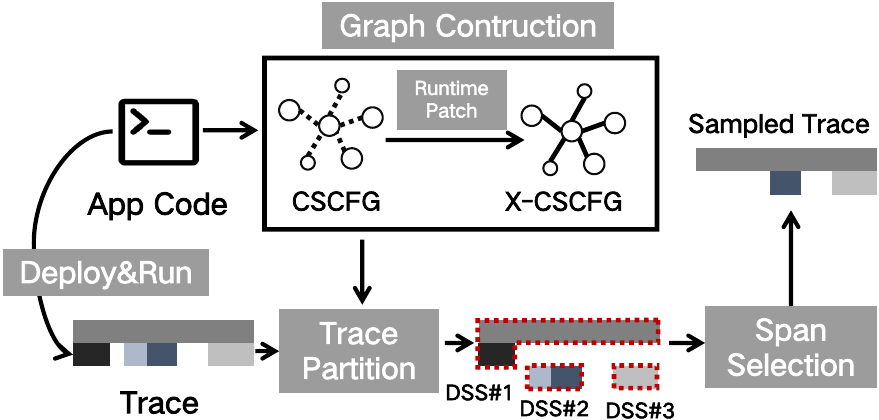} 
\caption{The Overview of AutoScope}
\label{fig:overview}
\end{figure}

To achieve Trace Sampling 2.0, we designed and implemented \method. As shown in Figure~\ref{fig:overview}, \method~ first performs static analysis to build a CSCFG, then refines it with dynamic information. During the sampling stage, it maps spans in the input trace to corresponding code functions and partitions the trace based on the CSCFG, thus identifying Dominate Span Set~(DSS). Finally, the system uses anomaly scores and sampling ratios to perform the final sampling at the DSS level.

\subsection{Trace Partition}

\subsubsection{CSCFG Construction}
To construct CSCFG, we first perform static analysis on the application code (\ie via Soot\footnote{\url{https://soot-oss.github.io/soot/}} and angr\footnote{\url{https://angr.io/}}), generating the application’s ICFG. We then traverse the basic blocks in the ICFG and retain only those containing application function calls, ultimately forming the CSCFG.

However, constructing the ICFG relies on interprocedural analysis, which is challenging under microservices architectures due to the prevalence of cross-service calls~\cite{ChenASE22, COCAICSE25}. These calls are made dynamically, often through network communication (\eg~gRPC or REST APIs), making it difficult to capture call edges between instances purely through static analysis~\cite{LiSecC22, SamhiICSE22}. Worse yet, the variety of frameworks used to implement these calls introduces further complexity, as each protocol and invocation mechanism differs. Existing approaches have largely left this issue unresolved~\cite{COCAICSE25, LiTX19}.

To address this, we propose an approach that patches the statically constructed CSCFG using runtime tracing information. While static analysis can infer certain cross-service calls, its effectiveness is mostly limited to explicit and direct invocations, such as the invocation in Figure~\ref{fig:BG_1}. While complex invocation mechanisms like Java reflection, widely used in frameworks like Spring Framework~\cite{SpringFramework} and gRPC, challenge static analysis. Our dynamic optimization targets these cases, bridging gaps where static methods fall short.

Specifically, we employ the wrk2 tool~\cite{wrk2} with enhanced workload generation capabilities to recover cross-service invocation relationships for the missing calls. While wrk2 excels at replaying industry-standard workloads and applying various load policies~\cite{yinfangAIOPs}, its native payload generation relies on predefined templates and certain fields, which may miss deep execution paths triggered by complex input combinations. To address this limitation, we augment wrk2's interface with LLM-powered payload generation specifically for services requiring nested parameter structures~\cite{KimSSSO24icse, AlmutawaGC24icdcs}. We aim to trigger exercise deeper service invocation chains, enabling both complex cross-service call discovery and production-like traffic simulation to collect comprehensive runtime tracing data. We refine the CSCFG from static analysis using runtime traces, bridging missed invocations through span call relationships.

\subsubsection{DSS Identification}

As shown in section~\ref{sec:CSCFG}, the dominance relationship between functions allows them to be inferable. Consequently, recording any single function within a group is sufficient to infer all functions in the same set. After constructing the CSCFG, we extend this dominance relationship to the corresponding spans to determine the dominance relationships among them. A group of spans that are mutually dominated is collectively referred to as the \textbf{Dominate Span Set (DSS)}. Formally, let $\{S_1, S_2, \ldots, S_n\}$ be a set of spans, and let their corresponding functions be $F_1, F_2, \ldots, F_n$. If for any $S_i$ and $S_j$ (where $i \neq j$), the functions $F_i$ and $F_j$ mutually dominate each other, then $\{S_1, S_2, \ldots, S_n\}$ is defined as a \textbf{DSS}.

Given the established dominance relationships, each DSS serves as a marker for branches, like the span set that indicates whether the seats are comfortable or economy in the motivating example in Figure~\ref{fig:motivation}. The presence of any span within a DSS indicates that the corresponding branch has been executed. Since spans within the same DSS are inferable from one another, retaining only one is sufficient to represent the execution path. Leveraging this property, Autoscope selects DSS as the fundamental unit for span-level sampling. The identification of DSS consists of three main steps:  

\textbf{Function Span Matching.} 
The first step is mapping spans to corresponding functions. In most cases, a span’s operation name follows the format \textit{Class.FunctionName}, while its metadata (e.g., \textit{pod name}) indicates the associated service package. By leveraging the service name, class name, and function name, we can accurately associate spans with their functions in the code.  However, certain cases may lead to mapping failures. For instance, user-defined library functions (e.g., \textit{ts-common} in the Trainticket system) are not tied to a specific service but are shared across multiple services. Spans generated by such functions may inherit metadata from the calling spans. Since the caller’s service does not actually contain the library function, this results in a mapping failure.  We build a dictionary specifically for user-defined functions that are not part of any single service to enhance the mapping accuracy.

\textbf{Execution Path Identification.}
In this step, our goal is to map traces to execution paths in the CSCFG. A straightforward approach would be to precompute and store all possible paths, then perform the longest sequence match during trace alignment. However, due to the path explosion problem in graphs generated by static analysis, storing all paths is computationally and storage-wise infeasible. Instead, we adopt an incremental path-matching strategy, dynamically traversing the CSCFG based on the function corresponding to each span to reconstruct the execution path.

While this method avoids the need for pre-stored paths, it presents another challenge: some spans fail to map to the CSCFG in earlier steps, preventing full coverage of all trace segments. This issue arises mainly because some spans do not correspond to concrete functions. For example, when OpenTelemetry monitors the Spring Web framework, URL mappings are often treated as independent spans without direct associations to specific functions, leading to missing functions in the graph.

To address these limitations, we dynamically adjust the CSCFG during path matching by inserting unmapped spans to fit the trace path. This ensures trace completeness by integrating unmapped spans, preventing loss of critical data, and enabling precise trace reconstruction afterward. We implement this with a dynamic programming approach and path edit distance to compute the optimal alignment, ensuring consistency with the CSCFG. To enhance efficiency, we introduce a caching mechanism that prioritizes lookups in cached trace mappings before directly traversing the graph.

\textbf{Trace Partition.}
After identifying the execution path of a trace within the CSCFG, we segment the trace based on the graph, dividing it into distinct DSSs. Specifically, when the path encounters a branch, we group the current forked span along with all preceding spans into the same DSS. As a result, a given trace can be decomposed into one or more DSSs. The selection of DSSs forms the foundation for maintaining trace structure. In subsequent analysis, we can simply map each span to its corresponding function and identify these functions on the CSCFG. This allows us to construct an execution path from the entry span to the leaf span. Using this reconstructed path, we can restore the trace’s span relationships and execution order, ultimately reconstructing its original structure.

\subsection{Branch Span Selection}

After obtaining the DSSs, we perform span selection based on them. This section explains how to select spans within these sets to achieve high-quality sampling. Specifically, we rank the spans using a revised Z-score and retain the top-k spans based on a user-defined sampling budget, ensuring an effective span-level sampling strategy.

\subsubsection{Z-score Calculation}
We first introduce how we rank the span with the Z-score~\cite{abdi2007z}. The Z-score quantifies how much a span deviates from historical performance, providing a measure of its anomaly level. However, span durations are often highly unstable with high variance~\cite{TennagePJJ19, EzazKE24, RahmanL19}. To address this, we use a variant of the Z-score model that relies on the median and median absolute deviation (MAD). This approach, combined with a dynamic sliding window mechanism, offers a more robust assessment of duration anomalies, particularly in scenarios with high variance~\cite{Cao18INFOCOM}.

The calculation formula is shown as \( Z_i = \frac{x_i - \text{median}(X)}{\text{MAD}} \), where \( X \) represents the set of durations for a given span within the time window, and \( x_i \) denotes the duration of the current span. Notably, the duration here refers to the actual execution time after excluding child spans.  The MAD (Median Absolute Deviation) is computed as \( \text{MAD} = \text{median} \left| x_i - \text{median}(x_i) \right| \), quantifies how much the durations in the current window deviate from the median. Given that span durations exhibit an unstable distribution, using the median and MAD offers greater robustness in anomaly detection compared to traditional Z-scores based on the mean and variance.  

To improve computational efficiency, we adopt a Min-Max Heap Pair~\cite{AtkinsonSSS86} to dynamically maintain the median and leverage the P² algorithm~\cite{jain1985p2} to estimate MAD. This approach enables incremental updates of statistical metrics within the sliding window while maintaining constant space complexity.

\subsubsection{Span Sampling}

After obtaining the weighted Z-score, we rank the spans within each Dominator Span Set (DSS) following the sampling algorithm described in Algorithm~\ref{alg:SpanSampling}. 

First, the sampling quota is proportionally allocated to each DSS based on its span count (lines 1-9), adhering to the user-defined budget. If a DSS receives a quota of less than one, we ensure at least one span is selected to maintain representation across all DSSs (lines 1–6). When the user-defined budget exceeds the minimum required to satisfy all DSSs, the remaining quota is distributed proportionally based on the number of spans within each DSS (lines 7–9). Consequently, in cases where the budget is limited, the final sampling ratio may exceed the predefined budget.

Next, we compute the Z-score for each span within a DSS and rank them accordingly (lines 10–12). However, high-ranking spans do not always indicate clear anomalies. If all Z-scores remain low, selecting the top-$k$ spans solely by rank offers little practical insight. To mitigate this, we introduce a threshold $\theta_z$, set at the 90th percentile, efficiently estimated using the P$^2$ algorithm with parabolic interpolation (line 13).

If the number of spans exceeding $\theta_z$ is insufficient to meet the allocated quota, we employ a Least Recently Sampled strategy. Specifically, spans are ranked based on their selection frequency within a given time window, prioritizing those sampled less frequently (lines 14–16). This improves coverage and ensures full utilization of the assigned quota. Ultimately, spans satisfying these criteria are selected from each DSS (line 17), completing the span-level sampling process.


\begin{algorithm}[t] 
    \caption{Sampling Dominant Span Sets}
    \label{alg:SpanSampling}
    \LinesNumbered
    \KwIn{DSS $D=\{D_1,\dots,D_n\}$, sampling ratio $p$, threshold $\theta_z$}
    \KwOut{Sampled spans $S$}

    $\text{totalBudget} \leftarrow \lfloor p \cdot \sum_{i=1}^{n} |D_i|\rfloor$

    \For{$i \leftarrow 1$ \KwTo $n$}{
      $B_i \leftarrow 1$
    }
    \uIf{$\text{totalBudget} < n$}{
      \For{$i \leftarrow 1$ \KwTo $n$}{
        $B_i \leftarrow 1$
      }
    }\Else{
      $\text{leftover} \leftarrow \text{totalBudget} - n$
      \For{$i \leftarrow 1$ \KwTo $n$}{
        $B_i \leftarrow 1 + \Big\lfloor \text{leftover} \times \dfrac{|D_i|}{\sum_{k=1}^{n} |D_k|} \Big\rfloor$
      }
    }

    $S \leftarrow \varnothing$
    \For{$i \leftarrow 1$ \KwTo $n$} {
      $Z \leftarrow \text{ComputeZScores}(D_i)$
      $G_i \leftarrow \{\, s \in D_i \mid Z(s) \ge \theta_z \,\}$
    
      $\text{tmpPick} \leftarrow \text{SelectTop}(G_i,\; B_i,\; Z)$
    
      \If{$|\text{tmpPick}| < B_i$}{
        $\text{rmd} \leftarrow B_i - |\text{tmpPick}|$
        
        $\text{tmpPick} \leftarrow \text{tmpPick} \;\cup\; \text{FillRemainder}(D_i,\; \text{rmd})$
      }
      $S \leftarrow S \;\cup\; \text{tmpPick}$
    }
    \Return $S$
\end{algorithm}

\section{Evaluation}

\subsection{Experiment Setup}

\subsubsection{Data Collection}
We evaluate the performance of \method~ on two widely used open-source microservice applications with a well-established experimental environment: Train ticket~\cite{TrainTicket} and Social Network~\cite{gan2019open}. Train ticket is a ticket booking system consisting of 41 microservices that communicate via REST APIs, while Social Network is built with C++ and thrift, both commonly used in previous research~\cite{yu2023nezha, lee2023eadro}. We collected end-to-end traces using OpenTelemetry~\cite{Opentelemetry} with Grafana Tempo~\cite{chakraborty2021grafana}. Specifically, to obtain traces of complete spans, we add span annotations to all functions in application code in both TrainTicket and Social Network. To validate the quality of sampled trace for downstream tasks, we introduced various typical performance degradations into randomly selected operations or services in both applications. Including resource faults, such as CPU contention and network delay, which are performed using ChaosBlade~\cite{ChaosBlade}, and code exceptions and errors returns were injected through code modifications, we set each fault duration to 3 minutes to emulate the process between fault occurrence to fix. Finally, we collected 33,255 and 12,421 traces in total for Trainticket and Social Network, respectively, with a problem-related trace ratio of 4.12\% on average for each dataset.

\subsubsection{Research Question}
We perform extensive experiments to validate the effectiveness of
\method~ via answering the following research questions.

\noindent\textbf{RQ1: To what extent \method~ reduce the trace size?} This research question investigates the effectiveness of \method~ in mitigating storage pressure for the directly generated traces. Unlike traditional sampling methods that allow users to specify a custom ratio, and sampling trace according to it. \method~ should first determine the DSS based on the CSCFG of the corresponding span execution. Due to the enforced selection constraints within DSS~(\ie at least one span in each set), \method~ introduces a \textit{Lowest Sampling Ratio} (LSR), which sets a lower bound on trace reduction. This experiment examines \method’s LSR across different datasets and explores its relationship with the underlying code structure and span size to assess its compression capability.

\noindent\textbf{RQ2: What is the quality of spans collected by \method~?}
To evaluate the quality of spans sampled by \method, we compare \method~ with different sampling methods to demonstrate its effectiveness. Specifically, we use the faulty span coverage as the metric. A higher coverage indicates better span quality.

Since the Lower Sampling Ratio (LSR) constraint exists, all sampling approaches maintain the same sampling ratio within each dataset (15\% for TrainTicket, 25\% for Social Network, both slightly above the LSR threshold). Firstly, we compare \method~ with commonly used trace-level sampling methods:
\begin{itemize}[leftmargin=*]

\item \textbf{Perch}~\cite{Pedro2018PERCH}: An offline sampling method using hierarchical clustering based on graph features. Perch groups trace using hierarchical clustering and select representative traces evenly from each group.

\item \textbf{Sifter}~\cite{Pedro2019sifter}: This method samples less common traces by maintaining a low-dimensional probabilistic model of common execution paths, and assigns higher sampling probabilities to traces with high prediction errors.

\item \textbf{TraceMesh}~\cite{Chen2024Tracemesh}: This method uses Locality Sensitive Hashing (LSH), and dynamically adjusts sampling decisions through clustering, enhancing the diversity of sampled traces.
\end{itemize}

Additionally, we select several span-level sampling methods to showcase \method's advantages, including random sampling and the Log\textsuperscript{2} method~\cite{ding2015log2}. The Log\textsuperscript{2} method detects anomalous code regions by comparing them with runtime and historical data. Given the similarity between code regions and spans, we include Log\textsuperscript{2} in comparison with \method~.

\noindent\textbf{RQ3: How do traces sampled by \method~ perform in downstream RCA?} 
We further evaluate the quality of sampled traces by assessing their effectiveness in the Root Cause Analysis (RCA) task. To do so, we apply the sampled traces to the SOTA automated RCA methods. Because Autoscope samples traces based on DSS, we can recover the trace structure using the sampled spans joint with CSCFG, where the missing duration is fulfilled by historical means and STDs, ultimately yielding a complete trace for root cause analysis, making them compatible with automated RCA and preserving the operational analysis chain. Specifically, we evaluate \method~ using the following RCA methods and compare its performance against the request-level sampling approach.

\begin{itemize}[leftmargin=*]
\item \textbf{TraceRCA}~\cite{li2021traceRCA}: Analyzes the ratio of normal to anomalous calls using association rules to identify the root cause service.

\item \textbf{TraceAnomaly}~\cite{li2020traceanomaly}: Detects anomalous traces by learning normal trace patterns offline and localizing the root cause online.

\item \textbf{MicroRank}~\cite{yu2021microrank}: Combines a personalized PageRank approach with spectral analysis to identify and rank root causes.

\item \textbf{TraceContrast}~\cite{zhangzhang}: Uses key paths from traces and applies contrastive sequence pattern mining and spectral analysis to pinpoint the root cause.
\end{itemize}

\noindent\textbf{RQ4: How efficient is \method~}
In this research question, we analyze the time overhead of \method. First, we compare its running time with sampling methods used in previous RQ to quantify its performance across different scenarios and evaluate overall efficiency. Then, we break down its components and measure the time cost at each stage to gain deeper insights into its performance characteristics.

\subsubsection{Evaluation Metrics}

To evaluate the performance of our proposed \method~ sampling model, we employ three commonly used metrics: \textit{sampling ratio}, \textit{Acc@N}, and \textit{MRR}.  

\textbf{Sampling Ratio.} Unlike conventional sampling methods, \method~ selects spans based on CSCFG and DSS, which imposes a minimum sampling ratio (\ie~$\#\text{sampled spans} / \#\text{all spans}$) rather than allowing an arbitrary sampling budget. The sampling ratio serves as an indicator of how well \method~ reduces trace storage overhead—a lower ratio signifies more efficient sampling.  

\textbf{Acc@N \& MRR.} To assess the quality of the sampled traces, we apply them to an automated RCA framework and measure their effectiveness using \textit{Acc@N} and \textit{MRR}. \textit{Acc@N} quantifies the proportion of correctly identified root causes found within the top-N (N = 1, 2, 3, ...) entities in the returned suspicious function list. The Mean Reciprocal Rank (MRR)~\cite{wiki:MRR} is a statistical metric that computes the average of the reciprocal ranks across multiple queries. Higher Acc@N and MRR values indicate better trace quality.

\section{EVALUATION RESULTS}\label{sec: evaluation}

\subsection{RQ1: Span Size Reducation}

In this research question, we evaluate the effectiveness of Autoscope’s sampling strategy in reducing the scale of traces by examining the LSR across two datasets. 

\begin{figure}[htbp]
    \centering
    \begin{minipage}{0.49\columnwidth} 
        \centering
        \includegraphics[width=\linewidth, keepaspectratio]{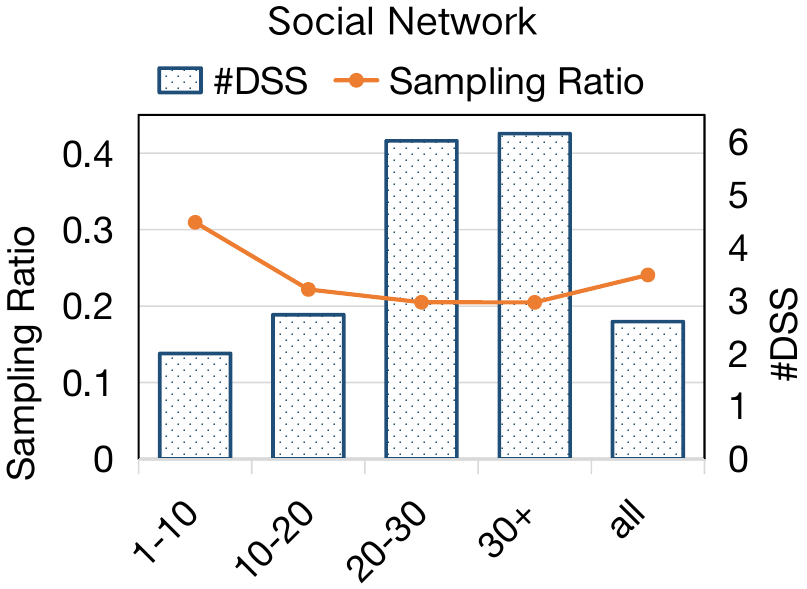}
    \end{minipage}
    \hfill 
    \begin{minipage}{0.49\columnwidth}
        \centering
        \includegraphics[width=\linewidth, keepaspectratio]{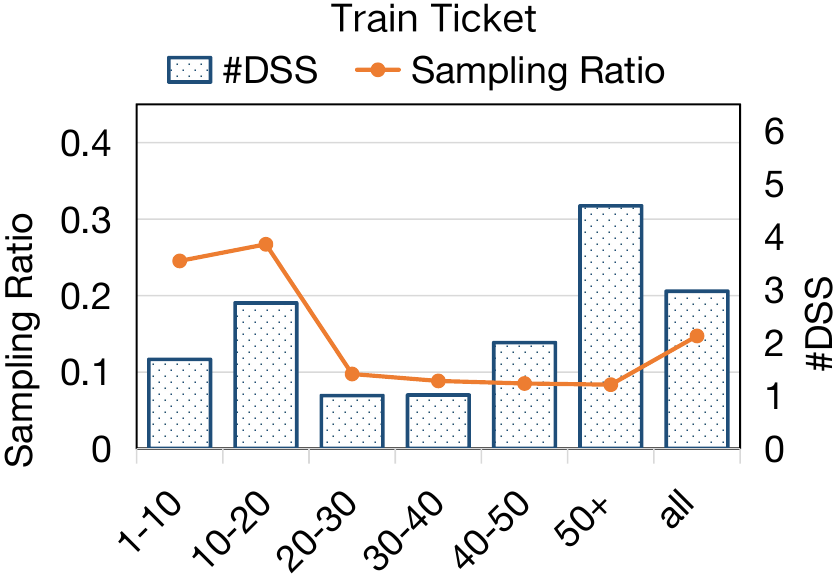}
    \end{minipage}
    \vspace{-4pt} 
    \caption{The sampling ratio with different span and DSS number across two datasets}
    \label{fig:RQ1}
\end{figure}

Figure~\ref{fig:RQ1} illustrates how the sampling rate changes with the number of spans in a trace and presents the average size of the \textit{DSS} across different span count intervals. As shown in the left part of the Figure, on the SN dataset, \method~ achieves the sampling rate of 24.9\% on average, and the value decreases as the number of spans increases, indicating improved sampling efficiency in traces with more spans. dropping from 30.9\% in the 1–10 span range to 20.4\% in the 30+ span range. This trend primarily arises because, while the number of DSS remains relatively stable, an increase in the number of spans leads to a higher average span count within each DSS. Since LSR selects only one span per DSS, a larger average span count within DSS results in a lower LSR. Although the number of DSS also increases with the span count, the growth is relatively small, rising from 1.9 in 1–10 span range to 6.1 in 30+ span range.

On the TT dataset, the sampling ratio exhibits a similar downward trend, averaging 14.7\%. However, an interesting anomaly emerges: as the span range increases from 1–10 to 10–20, the sampling ratio rises from 24.5\% to 26.7\%. This increase occurs because, despite the higher span count, the average number of DSS also grows from 2.0 to 2.7, requiring more spans to be selected and thus raising the sampling ratio. This observation suggests that while the sampling ratio is influenced by span count, it is also shaped by the branching structure of the code (\ie~ DSS).

\vspace{2mm}
\begin{center}
\find{
    \textbf{Answer to RQ1}: \method~ effectively samples traces while retaining all request records. It achieves an average sampling rate of 19.8\% across two datasets and performs better on traces with more spans.
}
\end{center}

\subsection{RQ2: Span Quality}

\begin{table}[h]
    \centering
    \caption{Faulty Span Coverage of Different Sampling Methods}
    \resizebox{\columnwidth}{!}{ 
    \begin{tabular}{lcccccc}
        \toprule
        \textbf{Dataset} & \textbf{Perch} & \textbf{Sifter} & \textbf{TraceMesh} & \textbf{Uniform} & \textbf{Log$^2$} & \textbf{Autoscope} \\
        \midrule
        TT(15\%) & 85.4 & 82.5 & 89.7 & 15.2 & 85.3 & 96.8 \\
        SN(25\%) & 89.7 & 88.7 & 94.8 & 24.7 & 91.1 & 99.3 \\
        Average  & 87.6 & 85.6 & 92.3 & 20.0 & 88.2 & 98.1 \\
        \bottomrule
    \end{tabular}
    }
    \label{tab:RQ2}
\end{table}

Table~\ref{tab:RQ2} presents the faulty span coverage of \method~compared to several trace-level and span-level sampling methods across two datasets. To ensure fairness, all methods are evaluated under the same sampling ratio. Since AutoScope imposes a minimum sampling rate constraint, we set the sampling ratios to 15\% and 25\% for the two datasets, both slightly above the threshold.

The results demonstrate that \method~ performs well on both datasets, achieving an average coverage of 98.1\%, ranking first. While TraceMesh, which employs multi-dimensional feature clustering, also achieves coverage above 90\%, it requires fault-free data for initial training. This requirement makes data collection costly especially in real-world production environments, whereas \method~does not have this limitation. Additionally, the trace-level sampling methods, Perch and Sifter, exhibited lower average coverage than \method. Log², which focuses solely on latency anomalies, did not demonstrate advantages, with coverage of 88.2\%.

For different types of faulty spans, \method~achieved an average coverage of 98.4\% for latency anomalies, surpassing Log\textsuperscript{2}, which is specifically designed to detect such anomalies. \method~enhances robustness in span-level sampling by excluding sub-span delays, considering only the delay of the current span, and applying a median-based Z-score for span selection. Regarding structural anomalies, \method~ effectively captures faulty spans due to its integration of code-level knowledge, particularly for conditional branches. For instance, in the \texttt{getToken} function of \textit{auth-service} in Train Ticket, an invalid ID triggers an early return, expressed as \texttt{if (!id) \{ return new Response<>(0, "Verification failed.", null); \}}. This results in traces with abnormal structures. \method~identifies control flow branches and generates corresponding DSS to select critical spans. Since DSS marks the abnormal branch, \method~naturally captures structural anomalies.

\begin{center}
\find{
    \textbf{Answer to RQ2}: \method~ achieves a high faulty span coverage across both datasets~(98.1\%), effectively capturing various types of faulty spans.
}
\end{center}

\subsection{RQ3: Downstream Analysis}


\begin{table}[h]
    \centering
    \caption{RCA Performance with Different Sampling Methods}
    \vspace{-5pt}
    \renewcommand{\arraystretch}{1.2}
    \begin{tabular}{@{}l l c c c c@{}}
        \toprule
        \textbf{RCA} & \textbf{Sampling} & \textbf{Acc@1} & \textbf{Acc@2} & \textbf{Acc@3} & \textbf{MRR} \\
        \midrule
        \multirow{5}{*}{TraceRCA} 
            & Perch     & 41.1 & 51.6 & 70.6 & 0.574 \\
            & Sifter    & 43.7 & 54.8 & 74.1 & 0.599 \\
            & TraceMesh & 51.9 & 62.4 & 78.7 & 0.661 \\
            & \textbf{AS w/o $\mathcal{C}$}& 28.2 & 47.1& 63.6& 0.522 \\
            & \textbf{AS} & \textbf{57.9} & \textbf{72.7} & \textbf{88.3} & \textbf{0.715} \\
        \midrule
        \multirow{5}{*}{\makecell{Trace\\Anomaly}} 
            & Perch     & 54.4 & 60.77 & 76.1 & 0.670 \\
            & Sifter    & 55.1 & 63.9 & 77.2 & 0.688 \\
            & TraceMesh & 62.1 & 69.8 & 81.0 & 0.734 \\
            & \textbf{AS w/o $\mathcal{C}$}& 31.4 & 48.3 & 62.3&  0.536 \\
            & \textbf{AS} & \textbf{71.9} & \textbf{79.9} & \textbf{91.1} & \textbf{0.812} \\
        \midrule
        \multirow{5}{*}{\makecell{Trace\\Contrast}} 
            & Perch     & 40.9 & 52.9 & 71.1 & 0.573 \\
            & Sifter    & 45.3 & 56.4 & 75.1 & 0.609 \\
            & TraceMesh & 54.1 & 66.3 & 80.7 & 0.678 \\
            & \textbf{AS w/o $\mathcal{C}$} & 25.6 & 41.2 & 68.4& 0.518\\
            & \textbf{AS} & \textbf{56.3} & \textbf{76.3} & \textbf{89.9} & \textbf{0.715} \\
        \midrule
        \multirow{5}{*}{MicroRank}
            & Perch     & 42.0 & 52.4 & 70.2 & 0.580 \\
            & Sifter    & 43.6 & 54.9 & 71.2 & 0.595 \\
            & TraceMesh & 51.4 & 62.9 & 78.1 & 0.658 \\
            & \textbf{AS w/o $\mathcal{C}$}& 24.4& 45.8 & 65.2& 0.513  \\
            & \textbf{AS} & \textbf{57.2} & \textbf{70.5} & \textbf{88.0} & \textbf{0.716} \\
        \bottomrule
    \end{tabular}
    \vspace{-5pt}
    \label{tab:rca_comparison}
\end{table}

In this research question, we evaluate sampling quality by assessing how different sampling strategies perform in different automated RCA approaches. The experimental results are shown in Table~\ref{tab:rca_comparison}, demonstrating that Autoscope sampling improves MRR performance across four RCA methods by 8.1\%, 10.6\%, 5.5\%, and 8.8\%, respectively. This suggests that span-level sampling for all traces with Autoscope significantly enhances downstream analysis. Among all methods, TraceAnomaly shows the most significant improvement, it relies on a VAE-based model for anomaly detection and root cause localization, which demands higher data quality and quantity than other RCA methods. By selecting key spans and reconstructing traces at the request level, Autoscope provides higher-quality training data for VAE. Additionally, for RCA methods like TraceContrast and MicroRank, which depend on comparative trace analysis, Autoscope enhances structural completeness by reconstructing each request’s trace, improving contrastive analysis.  Additionally, we conduct an ablation study to evaluate span-level sampling without code knowledge support. As shown in the table, this approach performs even worse than the weakest trace-level sampling strategy, achieving only 27.4\% acc@1 and 0.52 MRR on average. This decline occurs because, without the CSCFG structure, the trace loses its original form, making the RCA method ineffective. 

A key advantage of Autoscope is its CSCFG-based sampling, which enables end-to-end trace reconstruction. Missing span duration values during sampling are estimated using historical mean and variance, ensuring RCA methods remain effective in diagnosing faults. In contrast, purely latency-based span selection introduces trace ambiguity, complicating manual diagnosis for SREs and rendering contrastive or learning-based RCA methods ineffective.

\begin{center}
\find{
    \textbf{Answer to RQ3}:\method's CSCFG-based span selection ensures that span-level sampling remains compatible with automated RCA methods, the full reserved traces leading to improved downstream performance (8.3\% MRR increase on average).
}
\end{center}

\subsection{RQ4: Sampling Efficiency}

\begin{figure}[htbp]
    \centering
    \includegraphics[width=0.46\textwidth]{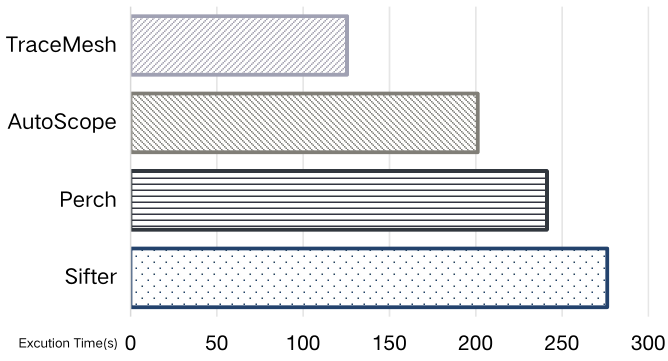}
    \caption{Time Cost between different sampling methods}
    \label{fig:Efficiency}
\vspace{-6pt}
\end{figure}

We evaluate the sampling efficiency of \method~ in this research question. The results are presented in Figure~\ref{fig:Efficiency}, which shows the total execution time of different methods on two datasets. \method~takes 201.1 seconds, approximately 4.4 ms per trace, which is comparable to Perch, which requires around 241.2 seconds. TraceMesh is the most efficient, completing the task in 125.4 seconds, while Sifter has the longest execution time at 276.1 seconds. During \method’s sampling process, trace partition accounts for 88\% of the computational cost, whereas span selection contributes only 12\%. Even though the graph construction time is excluded from the results as it is a one-time effort, the time dealing with the graph still contributes the most.

\begin{center}
\find{
    \textbf{Answer to RQ4}: \method~ maintains a high sampling quality while keeping computational overhead at a reasonable level. The primary cost lies in trace partition.
}
\end{center}



\section{Discussion}

\subsection{Practical Application}

Our proposed \method~ offers advantages to different stakeholders such as service providers and SREs. Its key impacts include:

\textbf{For Service Providers.}  Existing trace sampling methods employ a binary (1 or 0) strategy, which results in substantial trace loss. In contrast, \method~ retains all trace records while reducing storage overhead by 80\% (RQ1), enabling efficient and comprehensive trace data storage. Moreover, \method's span-based sampling strategy is orthogonal to existing request-based sampling approaches, which means they can be combined. For instance, request-level sampling can be applied first, followed by \method's span-level filtering, or vice versa, further minimizing storage costs. Although \method~ requires constructing a CSCFG for trace partition and recovery, this cost is one-time, with incremental updates ensuring sustainability and low maintenance even as the code evolves.  

\textbf{For SREs.} By preserving all trace records, \method~ provides more comprehensive data support for automated RCA, enhancing feature learning and comparative analysis. As demonstrated in the result of RQ2 and RQ3, various RCA methods obtain better performance under AutoScope-sampled data. Additionally, SREs can query traces for all requests, even those that do not exhibit anomalies—an essential capability for diagnosing complex problems such as off-the-path issues. 


\subsection{Threats to Validity}

The construction of CSCFG faces inherent limitations due to static analysis boundaries \cite{LiSecC22, Wang2024ISSTA, SamhiICSE22}, particularly in handling dynamic invocations (e.g., network interactions), programming language reflection mechanisms, and multi-threaded operations in distributed systems. These constraints may affect the completeness of control flow and invocation relationship representation. Further, hindering all CSCFG-based analysis.

To address these challenges, we implement two key mitigation strategies. First, the static CSCFG is enhanced with execution-based traces to dynamically patch gaps caused by unresolved dynamic bindings. Second, a specified load generator is employed to produce diverse and complex workloads, maximizing execution path coverage to improve trace collection completeness. This hybrid approach aims to reconstruct comprehensive execution flows in microservice systems by combining static analysis with empirical runtime observations. Together, these strategies mitigate the inherent deficiencies of purely static approaches, ensuring a more robust representation of execution behaviors for further analysis.

\section{Related Work}

\textbf{Trace Sampling approaches.}
With the exponential increase in trace volume in production systems, trace sampling has become a critical technique for managing data overload and ensuring system efficiency. Traditional tracing systems such as Dapper~\cite{sigelman2010dapper}, Jaeger~\cite{Jaeger}, and Zipkin~\cite{Zipkin} have employed uniform random sampling to mitigate storage overhead.
However, this approach fails to guarantee the representativeness and quality of the sampled traces, potentially leading to incomplete or misleading insights.
To address these limitations, recent studies have explored biased sampling techniques, leveraging methods such as tree-based models~\cite{huang2021sieve}, clustering algorithms~\cite{Pedro2018PERCH}, and neural language processing techniques~\cite{Pedro2019sifter}. These approaches primarily focus on identifying and preserving anomalous traces while minimizing the storage of normal traces.
For instance, STEAM~\cite{he2023steam} employs Graph Neural Networks (GNNs) to represent traces and sample mutually dissimilar traces, thereby enhancing system observability.
Similarly, Hindsight~\cite{zhang2023benefit} introduces the concept of retroactive sampling, which aims to capture traces of symptomatic edge cases retrospectively.
Despite their advantages, as discussed in Sec.~\ref{sec:why_we_need}, these trace-level sampling methods compromise trace queryability and fault diagnosis by disregarding entire normal traces.
To overcome this limitation, Astraea~\cite{toslali2024online}, the most closely related work to ours, proposes a span-level probabilistic sampling strategy that integrates online Bayesian learning and multi-armed bandit frameworks to assess the utility of spans and selectively discard them.
However, Astraea primarily considers duration variance while overlooking the structural information within traces, which may result in the ambiguity problem shown in section~\ref{sec:Motivating}.
In contrast, our method \method~ utilizes code information to preserve trace structure, ensuring consistency and compatibility with downstream tasks.

\noindent\textbf{Trace-based analysis approaches.} In distributed system performance diagnosis, traces play a critical role, serving as the foundation for numerous analyses. For instance, in anomaly detection, DeepTraLog~\cite{zhang2022deeptralog} integrates traces with logs into a graph structure and employs Graph Neural Networks (GNNs) for training, enabling cross-service anomaly detection Similarly, TraceCRL~\cite{zhang2022tracecrl}. leverages contrastive learning on operation-call graphs to obtain rich trace representations, significantly improving detection performance. Many automated trace-based RCA approaches have also been developed~\cite{li2021traceRCA, li2020traceanomaly, yu2021microrank}, with some studies~\cite{yu2023nezha, lee2023eadro} incorporating multidimensional information to enhance fault localization. These methods rely on high-quality trace data. However, traditional sampling strategies often result in significant trace loss, reducing analytical accuracy. In contrast, Autoscope adopts a span-level sampling approach that preserves all trace records while effectively minimizing storage overhead, ensuring high-quality data for trace-based analysis.

\section{Conclusion}

In this paper, we introduce the concept of Trace Sampling 2.0, a span-level sampling strategy that preserves structural integrity while reducing storage costs. To implement the concept, we design and develop Autoscope, which samples critical spans from traces while leveraging static analysis to extract execution logic from the application in the form of CSCFG, ensuring that the trace structure is retained. Evaluation results show that Autoscope achieves a high sampling ratio and quality, demonstrating its effectiveness in production environments.

\balance
\bibliographystyle{ACM-Reference-Format}
\bibliography{reference}

\end{document}